\documentclass[letterpaper]{article} % DO NOT CHANGE THIS
\usepackage{aaai25}  % DO NOT CHANGE THIS
\usepackage{times}  % DO NOT CHANGE THIS
\usepackage{helvet}  % DO NOT CHANGE THIS
\usepackage{courier}  % DO NOT CHANGE THIS
\usepackage[hyphens]{url}  % DO NOT CHANGE THIS
\usepackage{graphicx} % DO NOT CHANGE THIS
\usepackage{natbib}  % DO NOT CHANGE THIS AND DO NOT ADD ANY OPTIONS TO IT
\usepackage{caption} % DO NOT CHANGE THIS AND DO NOT ADD ANY OPTIONS TO IT
\usepackage{algorithm}
\usepackage{algorithmic}

\usepackage{newfloat}
\usepackage{listings}
\DeclareCaptionStyle{ruled}{labelfont=normalfont,labelsep=colon,strut=off} % DO NOT CHANGE THIS
\floatstyle{ruled}
\newfloat{listing}{tb}{lst}{}
\floatname{listing}{Listing}
\usepackage{xcolor}
\usepackage{placeins}

\nocopyright %-- Your paper will not be published if you use this command
\title{RAVEN: An Agentic Framework for Multimodal Entity Discovery from Large-Scale Video Collections}
\author {
    Kevin Dela Rosa
}
\affiliations{
    Aviary Inc.\\
    San Francisco, CA\\
    kdr@aviaryhq.com
}
\usepackage{bibentry}
\begin{document}

\maketitle
	
\begin{abstract}
	We present RAVEN (\textbf{R}ecognition and \textbf{A}daptation of \textbf{V}ideo \textbf{EN}tities), an adaptive AI agent framework designed for multimodal entity discovery and retrieval in large-scale video collections. Synthesizing information across visual, audio, and textual modalities, RAVEN autonomously processes video data to produce structured, actionable representations for downstream tasks. Key contributions include (1) a category understanding step to infer video themes and general-purpose entities, (2) a schema generation mechanism that dynamically defines domain-specific entities and attributes, and (3) a rich entity extraction process that leverages semantic retrieval and schema-guided prompting. RAVEN is designed to be model-agnostic, allowing the integration of different vision-language models (VLMs) and large language models (LLMs) based on application-specific requirements. This flexibility supports diverse applications in personalized search, content discovery, and scalable information retrieval, enabling practical applications across vast datasets.
\end{abstract}
	
\section{Introduction}
The exponential growth of video content across platforms necessitates intelligent systems for organizing and retrieving information at scale. Video collections, spanning domains such as education, entertainment, and instructional content, present unique challenges due to their multimodal nature—needing to integrate visual, auditory, and textual data.

Recent advances in large language models (LLMs) and vision-language models (VLMs) enable new opportunities for multimodal understanding \cite{zhang2023videollamainstructiontunedaudiovisuallanguage,maaz2024videochatgptdetailedvideounderstanding, lin2023videollavalearningunitedvisual}. However, these methods typically process videos in isolation  focusing on individual video comprehension, but lack mechanisms for collection-wide analysis. This capability is crucial for applications requiring a cohesive understanding of video collections rather than isolated clips.

RAVEN addresses these gaps with a model-agnostic design, allowing the integration of different VLMs and LLMs to suit domain-specific needs. This ensures adaptability to application-specific constraints such as computational efficiency or context length requirements. Our contributions include:

\begin{itemize}
	\item A modular agentic architecture for video category cannonicalization and multimodal entity extraction.
	\item A synthetic example and schema-guided mechanism for contextual prompting.
	\item Demonstration of high-quality structured entity extraction across large-scale video datasets using popular off the shelf LLM/VLMs.
\end{itemize}
	
\section{RAVEN Framework Overview}

\begin{figure*}
	\centering
	\includegraphics[width=0.8\linewidth]{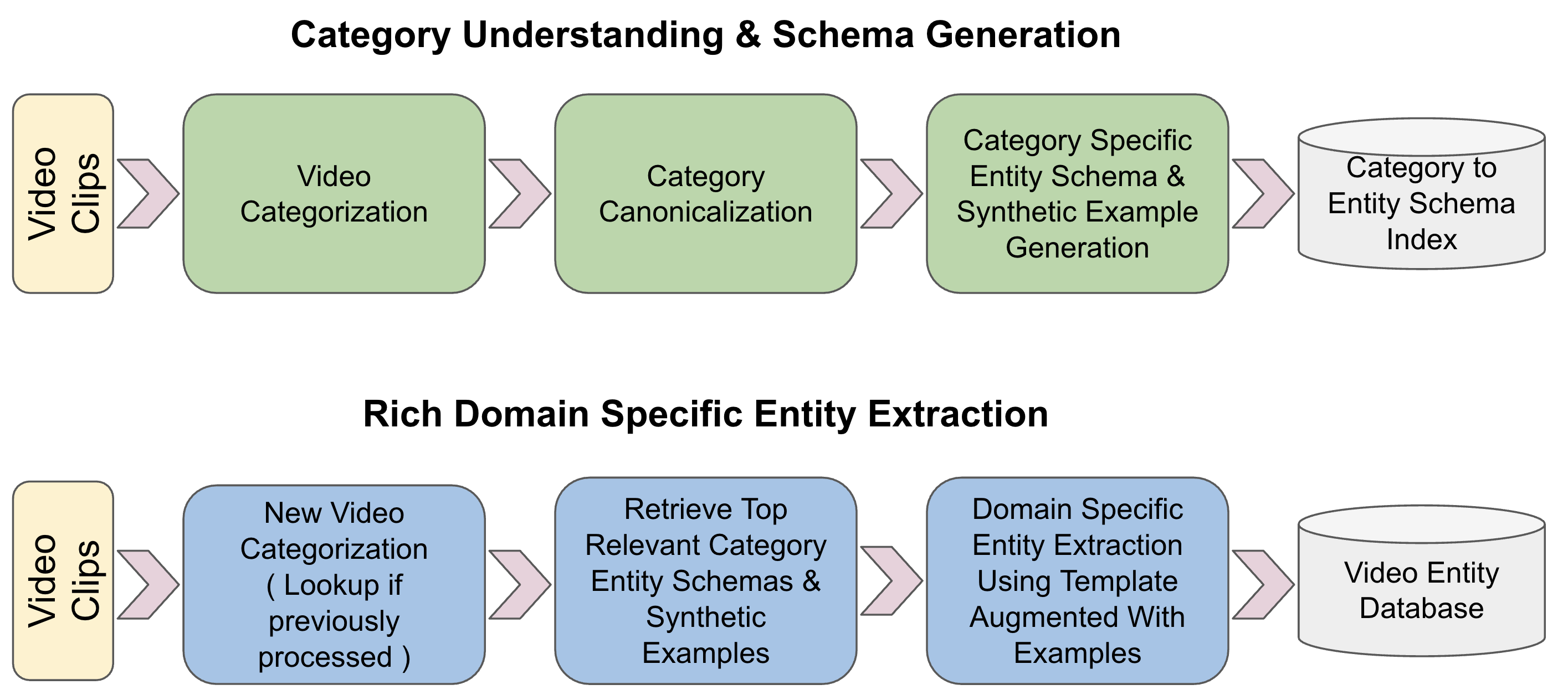}
	\caption{RAVEN Framework Overview. The two main LLM / VLM driven flows: 1) Category Understanding \& Schema Generation, 2) Rich Domain Specific Entity Extraction}
	\label{fig:system}
\end{figure*}

RAVEN operates as an adaptive agent for structuring multimodal video data, allowing one to operate on a collection of videos without necessarily having deep knowledge of the content contained within the collection but still allow for rich and consistent extraction of structured information across the provided videos. Our framework is comprised of two core stages: \textbf{Category Understanding}, and \textbf{Rich Domain Specific Entity Extraction}

Throughout these stages, as illustrated in Figure \ref{fig:system}, RAVEN uses a vision-language model for video-based tasks such as categorization and entity extraction, and a large language model for category name consolidation and generation. This frame work is designed to be model-agnostic as long as they handle comparable context lengths and support structured output JSON representations. For this study we used Gemini 1.5 Flash \citep{deepmind_gemini_flash} as our VLM (using both visual video and audio input) and GPT-4o \citep{openai2023gpt4} as our general text LLM.

\subsection{Category Understanding \&  Schema Generation}

Video clips are first processed by the VLM to infer \textit{video categories} and if desired extract \textit{general-purpose entities} such as people, objects, and locations in same prompt. This process can optionally include user-provided prompts to steer categorization toward specific goals. Then the top occurring raw category names produced by the VLM are fed through an LLM to normalize and dedupe similar concepts to produce a canonical list of categories.

Using the canonical categories, the LLM generates a list of typical entities expected in that domain and produces corresponding \textit{domain-specific entity schemas}. For each category, the generated schema includes:

\begin{itemize}
	\item A list of \textit{typical entities}.
	\item Attributes for each entity, with \textit{descriptions} and \textit{example} values.
\end{itemize}

The resulting entity schema lists per category are indexed by category name for later retrieval/lookup to help guide the actual entity extraction process. This modular design ensures adaptability for domain-specific needs, reducing manual intervention in schema development.

\subsection{Rich Domain Specific Entity Extraction}

Finally, video clips are processed again (as well as any desired additional clips) through the VLM. Using the assigned category from the first agentic flow, the system retrieves the most relevant schema based on \textit{semantic similarity} of the original unnormalized category to the top matching canonical category name. The schema is integrated into a prompt template with example values, facilitating in-context learning to extract entities and attributes aligned with the schema's structure. The results are then persisted and indexed according to the needs of the downstream application.

\section{Experiments \& Evaluation}

In this section we evaluate RAVEN's performance in its two core stages. We explore RAVEN's ability to infer video categories and generate schemas in Section \ref{sec:cu}, and in Section \ref{sec:er} investigate RAVEN's entity extraction capabilities.

\subsection{Category Understanding \& Schema Generation Analysis}
\label{sec:cu}

To demonstrate RAVEN's effectiveness in category understanding and schema generation, we applied RAVEN on 1.5 million video clips (over 5000 hours of footage) from the \textit{Aligned Video Captions} dataset \citep{rosa2024videoenrichedretrievalaugmented}, and report results from this large scale qualitative exploration. 

\begin{figure}
	\centering
	\includegraphics[width=0.9\columnwidth]{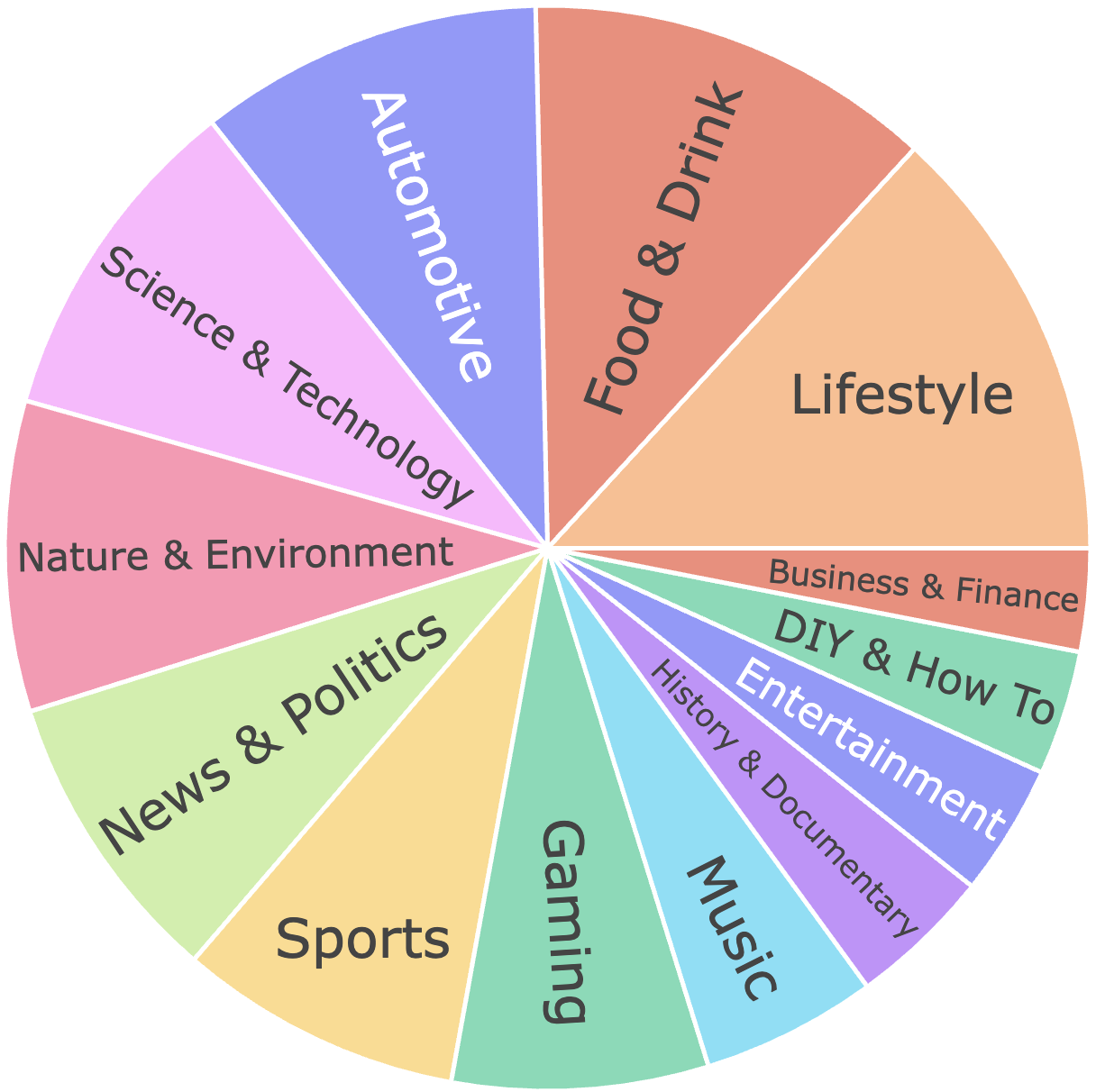}
	\caption{Inferred canonical video category distribution}
	\label{fig:qual1}
\end{figure}

In	Figure \ref{fig:qual1} we show the distribution of video clips by inferred canonical category, the individual categories and spread align nicely with the source dataset which was sampled equally from from YouTube's 15 top level categories.

\begin{figure}
	\centering
	\includegraphics[width=0.9\columnwidth]{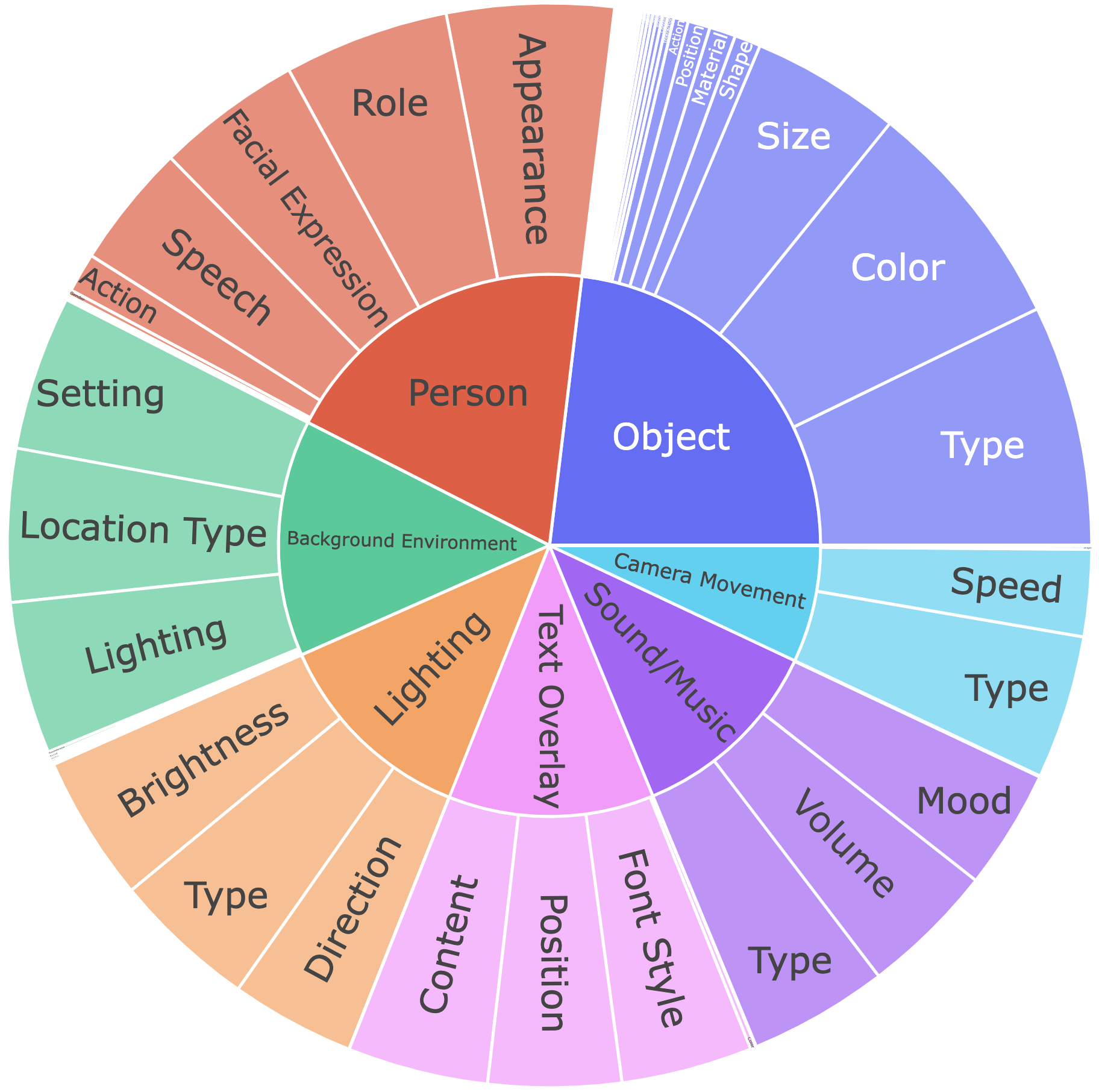}
	\caption{Distribution of extracted structured attributes for each generic entity type}
	\label{fig:qual2}
\end{figure}

Figure \ref{fig:qual2} presents the distribution of generic entity types and attributes, highlighting the flexibility of our framework to capture widely applicable generic entities with their associated attributes. 

\begin{figure}
	\centering
	\includegraphics[width=0.9\columnwidth]{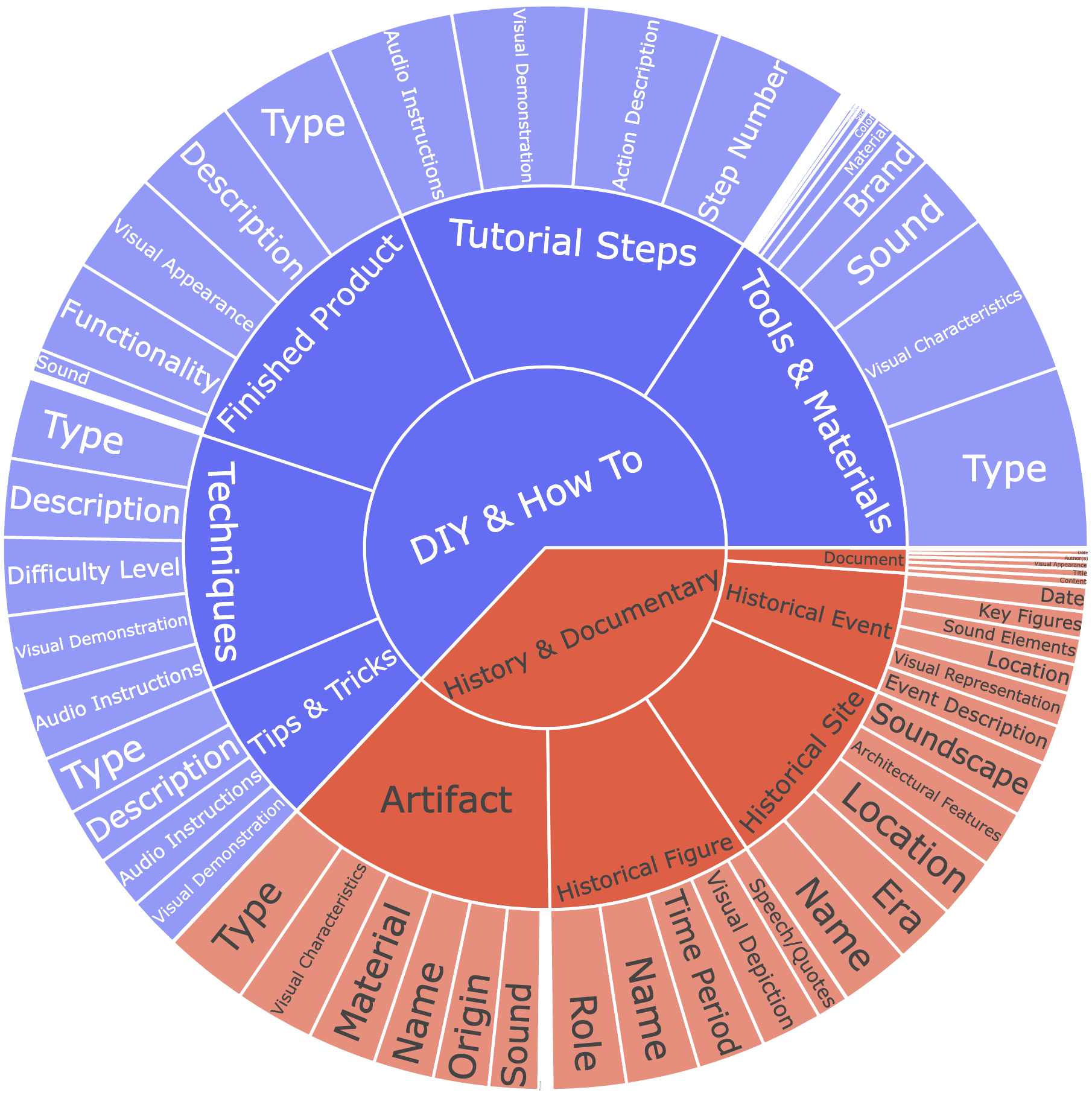}
	\caption{Distribution of domain-specific entity types \& attributes for How-To and History videos}
	\label{fig:qual3}
\end{figure}

In Figure \ref{fig:qual3}, we demonstrate the framework’s capacity for domain-specific extractions, exemplified by \textbf{How-To} and \textbf{History} videos, where we show the generated entity types and attributes, and their distributions in the dataset. Table \ref{tab:entities} lists the most frequently extracted entity value for sample entity types in different domains.

\begin{table}
	\centering
	\begin{tabular}{|p{1.2cm}|p{2.9cm}|p{3.1cm}|}
		\hline
		\textbf{Category} & \textbf{Entity $\rightarrow$ Attribute} & \textbf{Top Values} \\ \hline
		Generic           & Person $\rightarrow$ Role              & speaker, host, child, listener, narrator, chef \\ \hline
		Generic           & Background $\rightarrow$ Setting         & kitchen, living room, studio, office, city street \\ \hline
		History           & Event $\rightarrow$ Description        & world war ii, apollo 11, vietnam war \\ \hline
		History           & Figure $\rightarrow$ Name              & neil armstrong, abraham lincoln, adolf hitler \\ \hline
		How-To            & Tools \& Materials $\rightarrow$ Type  & knife, pot, bowl, camera, fishing rod \\ \hline
		How-To            & Techniques $\rightarrow$ Type          & cutting, cooking, mixing, installation \\ \hline
		Travel            & Destination $\rightarrow$ Location     & bangkok, new york city, tokyo, london \\ \hline
	\end{tabular}
	\caption{Sample entity values extracted for generic and domain-specific entities, showcasing RAVEN’s versatility and domain adaptation}
	\label{tab:entities}
\end{table}

\begin{table*}[tb]
\centering
\caption{Case study of extracted entities by method. This table illustrates the qualitative differences in entity extraction across methods for a sample video (e.g., a historical documentary about Abraham Lincoln )}
\label{fig:case-study}
\begin{tabular}{ p{5cm} | p{4cm} | p{2cm}|  p{1.9cm} | p{1.5cm} | p{1cm} }
	\hline
	\textbf{Entity → Attribute} & \textbf{Ours} & \textbf{Speech} & \textbf{OCR} & \textbf{Caption} & \textbf{YOLO} \\
	\hline
	\multicolumn{6}{ c }{\textbf{Class Agnostic Generic Entity}} \\
	\hline
	\textbf{Person → Role}       & Abraham Lincoln $\rightarrow$ President & Abraham Lincoln; President Lincoln & PRESIDENT LINCOLN & Abraham Lincoln & Person \\
	\textbf{Person → Gender}       &  Male & - & - & Man & - \\
	\textbf{Person → Age}       &  Mid 50s & - & - & - & - \\
	\textbf{Person → Appearance}       &  Wearing a Dark Suit & - & - & - & - \\
	\textbf{Person → Mood}       &  Sad Reflective & - & - & - & - \\
	\hline
	\textbf{Object → Type}       & Train Car & Train & - & Train Car & Train \\
	\textbf{Object → Color}       & Black \& White & - & - & - & - \\
	\textbf{Object → Size}       & Large & - & - & - & - \\
	\hline
	\multicolumn{6}{ c }{\textbf{History \& Documentary Specific Entities}} \\
	\hline
	\textbf{Historical Event → Description}       &  Surrender of the Army of Northern Virginia & Battle of Appomattox courthouse & - & - & - \\
	\textbf{Historical Event → Date}       &  April 9, 1865 & - & - & - & - \\
	\textbf{Historical Event → Location}       &  Appomattox Courthouse, Virginia & - & - & - & - \\
	\textbf{Historical Event → Key Figures}       &  Robert E. Lee, Ulysses S. Grant & - & - & - & - \\
	\hline
	\textbf{Historical Site → Location}       & Lincoln Memorial $\rightarrow$ Washington, D. C. &  Lincoln Memorial & - & - & - \\
	\textbf{Historical Site → Era}       & Early 20th Century & - & - & - & - \\
	\textbf{Historical Site → Architectural Features}       & Marble structure, neoclassical design & - & - & - & - \\
	\hline
\end{tabular}
\end{table*}

\subsection{Domain Specific Entity Richness Analysis}
\label{sec:er}

\begin{figure}[t]
	\centering
	\includegraphics[width=0.9\columnwidth]{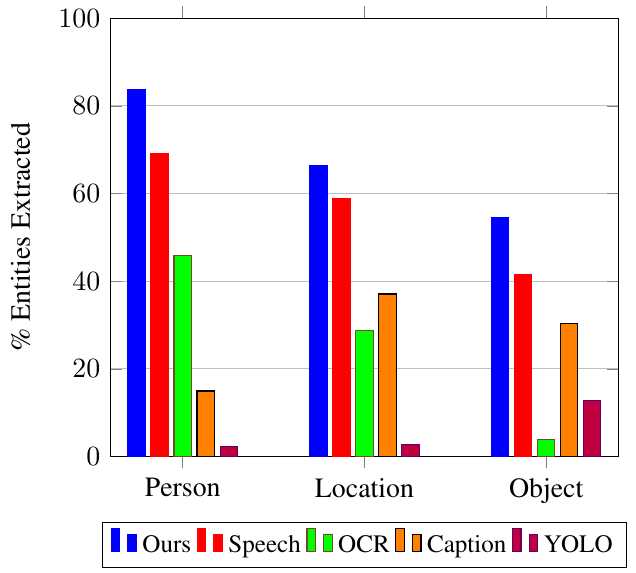}
	\caption{Entity Recall by method. Each bar represents the proportion of entities extracted by each method for the Person, Location, and Object entity types respectively}
	\label{fig:entity-recall}
\end{figure}

For understanding the quality of entities extracted in the framework, we evaluated RAVEN on a sample of 300 video clips ($\sim$10 seconds each) drawn from the \textit{Aligned Video Captions} dataset \cite{rosa2024videoenrichedretrievalaugmented}. The clips were selected from diverse categories, including \textit{Travel}, \textit{Education}, \textit{History}, and \textit{Instructional Content}, to ensure broad domain coverage. Each clip was processed through the RAVEN pipeline, starting with category inference and schema generation, followed by rich entity extraction.

To benchmark performance, we compared RAVEN against the following baselines using standard configurations:
\begin{itemize}
	\item \textbf{NER on Speech:} Identifies entities from transcribed video speech, extracted automatically via AssemblyAI Speech-to-Text \citep{assemblyai2023}. The named entities were extracted from the speech transcript using GLiNER \citep{zaratiana2023gliner}.
	\item \textbf{OCR on Scene Text:} Extracts visible text from frames, capturing entities like place names, using EasyOCR \citep{easyocr2020}.
	\item \textbf{Keyword Extraction from Visual Caption} Extracts keywords from the visual captions provided in Panda-70M \citep{chen2024panda70m}.
	\item \textbf{YOLO Object Detection:} Detects general objects in frames, labeling them without contextual structure, using YOLOv10 \citep{wang2024yolov10} pretrained on COCO dataset \citep{cocodataset}.
\end{itemize}

The baseline methods were selected to evalate different aspects of entity extraction. NER on Speech assesses linguistic entity recognition, OCR captures text from visual data, and visual captioning provides descriptive context. These baselines highlight specific extraction limitations that our framework addresses comprehensively by operating in a multimodal fashion.

Figure \ref{fig:entity-recall} visualizes the ability of each method to extract class agnostic generic entities from various videos. Our framework shows a strong ability to extract  \textit{Person}, \textit{Location}, and \textit{Object} entities. The baselines often lacked multi-modal context and finer grained vocabulary to successfully extract entities presented in the videos. Figure \ref{fig:entity-recall} demonstrates RAVEN's ability to synthesize multimodal context, significantly improving recall rates over unimodal baselines

Futhermore, we present a case study on a historical documentary to illustrate the qualitative depth of extracted entities, for both the class agnostic generic setting and the domain specific setting. Table \ref{fig:case-study} shows that our framework not only extracts named entities but also attributes, descriptions, and relationships (e.g., \textit{Person → Role, Event → Location}). Baseline methods produce isolated labels limiting their utility in structured retrieval. The qualitative analysis (Table \ref{fig:case-study}) demonstrates RAVEN's capability to extract fine-grained, contextual attributes compared to baseline methods, which often fail to capture relationships and attribute.

\section{Conclusion}
We presented RAVEN, an agentic AI framework for multimodal entity discovery and retrieval. By integrating category understanding, schema generation, and retrieval augmented \& example-guided extraction, RAVEN addresses the challenges of structuring unstructured video content. RAVEN demonstrates the ability to extract structured, domain-specific representations, advancing multimodal information retrieval. Future work will explore extending RAVEN's capabilities to support entity relationships and optimizations for schema generation \& signal extraction process. RAVEN's scalability and modularity position it as a versatile solution for evolving multimodal retrieval challenges.

\FloatBarrier

\bibliography{document}

\end{document}